\documentstyle[epsfig, twocolumn]{iso98}
\newcommand{\thepapername}{}

  \def\mic{\,\mu\rm m} 
\def\fu{{\rm W}/{\rm cm}^2/\mic}
\def\Teff{T_{\rm eff}}      

\def\HH{H{\sc ii}}	
\def\arii{[Ar\sc ii\rm]}   \def\ariii{[Ar\sc iii\rm]}   
\def\neii{[Ne\sc ii\rm]}   \def\neiii{[Ne\sc iii\rm]}   
\def\siii{[S\sc iii\rm]}   \def\siv{[S\sc iv\rm]}
\def\brgam{Br-$\gamma$}    \def\paa{Pa-$\alpha$}

\def\Lsun{L_\odot} 	   
\def\Av{A_{\rm V}}	   \def\Ak{A_{\rm K}}
     \def\tausil{\tau_{\rm sil}}

\def\lapp{\;\raise.4ex\hbox{$<$}\kern-0.8em \lower .6ex\hbox{$\sim$}\;}
\def\gapp{\;\raise.4ex\hbox{$>$}\kern-0.8em \lower .6ex\hbox{$\sim$}\;}

\begin{document}

\setlength{\parindent}{0pt}
\setlength{\parskip}{ 10pt plus 1pt minus 1pt}
\setlength{\hoffset}{-1.5truecm}
\setlength{\textwidth}{ 17.1truecm }
\setlength{\columnsep}{1truecm }
\setlength{\columnseprule}{0pt}
\setlength{\headheight}{12pt}
\setlength{\headsep}{20pt}
\pagestyle{veniceheadings}

\title{
\bf SPECTROSCOPY OF THE PISTOL AND QUINTUPLET STARS \\ IN THE GALACTIC
CENTRE\thanks{ISO is an ESA project with instruments funded by ESA Member
States (especially the PI countries: France, Germany, the Netherlands and
the United Kingdom) and with the participation of ISAS and NASA.}
}

\author{{\bf A.~Moneti$^{1,2}$, J.A.D.L.~Blommaert$^1$, F.~~Najarro$^3$,
D.~Figer$^4$, S.~Stolovy$^4$}  \vspace{2mm} \\
$^1$ISO Data Centre, ESA Astrophysics Division, Villafranca del Castillo,
Spain.\\
$^2$On contract from SERCo F.M.~B.V. \\
$^3$IEM, Consejo Superior de Investigaciones Cientificas, Madrid, Spain. \\
$^4$University of California, Los Angeles, CA, USA.\\
$^5$Steward Observatory, Tucson, AZ, USA.\\
}

\maketitle

\begin{abstract}

We present initial results of a spectroscopic study of the Pistol and of
the cocoon stars in the Quintuplet Cluster.  From ISOCAM CVF 5---$17\mic$
spectroscopy of the field of the Pistol Star, we have discovered a nearly
spherical shell of hot dust surrounding this star, a probable LBV.  This
shell is most prominent at $\lambda \gapp 12\mic$, and its morphology
clearly indicates that the shell is stellar ejecta.  Emission line images
show that most of the ionised material is along the northern border of this
shell, and its morphology is very similar to that of the Pistol \HH\ region
(Yusef-Zadeh \&\ Morris, 1987).  We thus confirm that the ionisation comes
from very hot stars in the core of the Quintuplet Cluster.  An SWS spectrum
of the Pistol Nebula indicates a harder ionising radiation than could be
provided by the Pistol Star, but which is consistent with ionisation from
Wolf-Rayet stars in the Quintuplet Cluster.  The CVF 5---$17\mic$ spectra
of the cocoon stars in the Quintuplet do not show any emission feature that
could help elucidate their nature.
\vspace {5pt} \\

Key~words: ISO, Galactic Centre, Quintuplet, Pistol, LBVs

\end{abstract}

\section{INTRODUCTION}

The Quintuplet Cluster is one of the few young clusters known in the
vicinity of the Galactic Centre.  It took its name from five very luminous,
$\sim 10^5\,\Lsun$ sources with very cool, 600---1200 K, energy
distributions (e.g.~Moneti, Glass, and Moorwood 1994, MGM94), henceforth
the \it cocoon\/ \rm stars, and it is now known to contain several dozen
hot stars (Figer, McLean, \&\ Morris 1999, FMM99, and references therein
for a detailed background) including several Wolf-Rayet stars, several OB
supergiants, and at least one and probably two luminous blue variables
(LBV; Figer et al.~1998, and references therein).  One of these is located
about 20 arcsec south of the cluster core and is now known as the \it
Pistol\/ \rm star because of its location at the center of curvature of the
G0.15--0.05 \HH\ region (Yusef-Zadeh \&\ Morris 1987), known as the \it
Pistol\/ \rm because of its shape, and itself located half way between the
cluster core and the Pistol star.  The Pistol \HH\ region was also detected
in \brgam\ (MGM), and was imaged in \arii\ (Nagata et al.~1996) and in
\paa\ (Figer et al.~1999), confirming its thermal origin (as opposed to the
non-thermal origin of the radio filaments which appear to cross it,
Yusef-Zadeh \&\ Morris 1987).

Early studies of the Pistol star indicated it had extremely high luminosity
($2\times 10^6\Lsun$, Figer, McLean \&\ Morris 1995, FMM95, and $10^7\Lsun$
Cotera et al.~1996).  There have also been suggestions that the Pistol Star
could be photoionising the Pistol \HH\ region (e.g.~Cotera et al.~1996),
while FMM95 suggested that the \HH\ region is material ejected from the LBV
which is photoionsed by the hot stars in the Quintuplet Cluster.  By
modeling its IR spectral energy distribution (SED) and its IR line
profiles, Figer et al.~(1998) constrained the Pistol Star's luminosity to
two classes of models, one with $L_* \approx 4\times 10^6\,\Lsun$ and
$\Teff \approx 14,000\,$K, and the second with $L_* \approx 15\times
10^6\,\Lsun$ and $\Teff \approx 21,000\,$K.  What remained unclear was why
the ejecta was only seen to the north of the star.

The nature of the cocoon stars remains mysterious.  Okuda et al.~1990 and
MGM94 suggested they might be star-forming cocoons similar to BN or
AFGL2591.  FMM99 favour extremely dusty WC stars, but also suggests that
they could form a new class of objects.  Glass et al.~(1999) report slight
variability of two of the five cocoon stars at $2.2\mic$, possibly
inclining the balance toward the evolved nature of these stars.  To date,
no emission feature has been detected toward these stars that could help
elucidate their nature, and the only absorption features are due to
silicates and ices (Nagata et al.~1996), all of which are believed to be of
interstellar origin.

\section{OBSERVATIONS AND RESULTS}

\begin{figure*}[t]  \begin{center} \leavevmode
\centerline{\epsfig{file=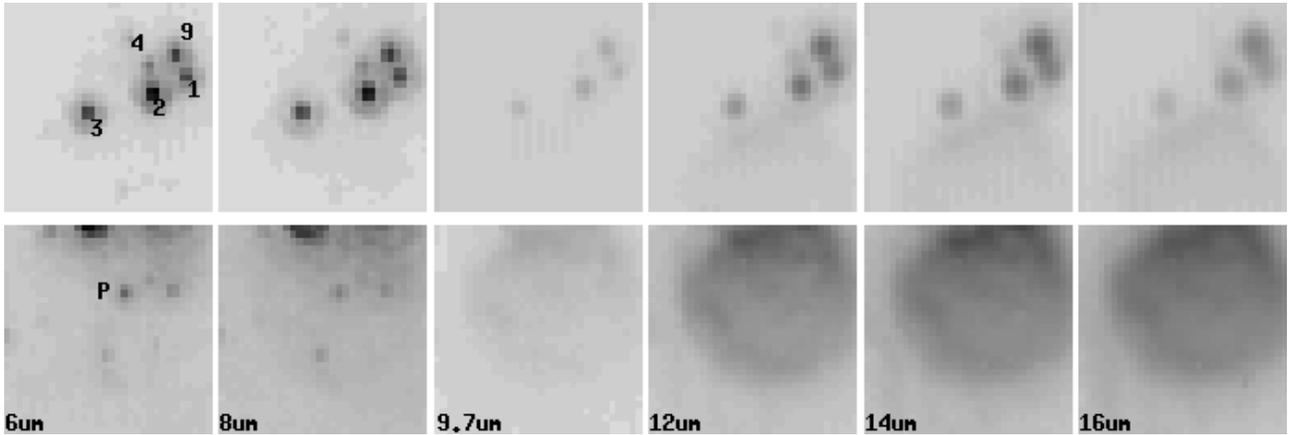,width=17cm}}
\end{center} \caption{\em Continuum images of the Quintuplet Cluster
(top) and of the Pistol Nebula (bottom). North is up, and east is to the
left.  The vertical stripes in the Quintuplet images are due to incomplete
dark subtraction at the 0.28 sec on-chip integration time.  Wavelengths
corresponding to each frame are marked, and the Quintuplet and Pistol stars
are identified.}
\end{figure*}

We have obtained ISOCAM-CVF scans of the Pistol and of the Quintuplet
fields, covering the 5.0---$17.1\mic$ spectral range.  All scans were
obtained with a pixel FOV of 1.5 arcsec, and the integration time was $\sim
30\,$sec per CVF position, using the 2.1 and the 0.28 sec on-chip
integration time on the Pistol and on the Quintuplet, respectively.  Data
reduction was performed with the CAM Interactive Analysis (CIA)
package\footnote{CIA is a joint development by the ESA Astrophysics
Division and the ISOCAM consortium, led by the ISOCAM PI, C.~Cesarsky,
Direction des Sciences de la Mati\`ere, C.E.A., France}, and followed the
standard steps of (i) dark subtraction, (ii) removal of cosmic ray events,
(iii) averaging all valid images obtained at a given CVF position, (iv)
flatfielding the results using library flats, (v) converting the average
signal to physical units using standard conversion factors, originally
derived from several standard stars.  No transient correction was applied
to the data presented here; given the brightness of the sources, the
transient correction is very minor, below a few percent, and for the
discussion presented here these deviations are not important.

Figure 1 shows a composite of continuum images of the Quintuplet and of the
Pistol.  They are scaled so that the intensities are proportional to the
observed energy density, $\lambda F(\lambda)$, with the scaling for the
Pistol being $5\times$ stronger than for the brighter Quintuplet Stars.
Both datasets clearly reveal the deep $9.7\mic$ silicate absorption
feature.  The images of the Quintuplet Cluster are dominated by the cocoon
stars, whose identification, following MGM94, is shown.  Since images are
diffraction limited, the FWHM of the images can be seen to increase with
wavelength, and the first Airy ring can also be recognised.  Also, the
brightest stars at short wavelengths are no longer so at long wavelengths,
showing their different colour temperatures.  The Pistol Star is the point
source located just above centre in the Pistol field, marked with a ``P''.
It is most clearly visible at short wavelengths, where other stars are also
detected.  The peak at the top edge of the frame corresponds to star no.~3
in the Quintuplet.  Beyond the silicate feature the Pistol Star is lost in
a nearly spherical centered on the Pistol star itself.  The diameter of
this Pistol Nebula is $\sim 1.5\,$pc.

\begin{figure}[bh]  \begin{center} \leavevmode
\centerline{\epsfig{file=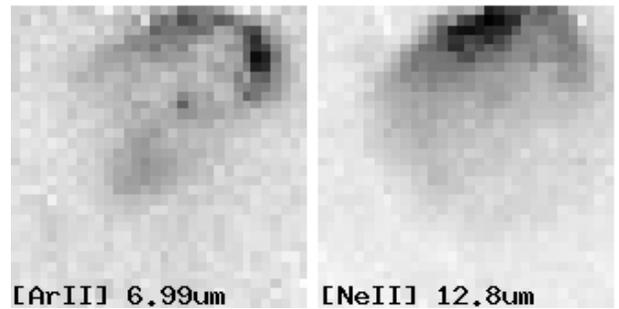,width=8cm}}  \end{center}
\caption{\em Line images of the Pistol Nebula. }
\end{figure}

Raw spectra extracted from the Pistol data cube show PAH emission at 6.2,
7.7, and 8.6 and $11.3\mic$, and the fine structure lines of \arii\
$7.0\mic$, \neii\ $12.8\mic$, and \neiii\ $15.4\mic$.  Line images reveal
that the PAH emission is uniform over the field, thus indicating that it is
interstellar rather than local to the nebula, while the forbidden line
emission is not.  The spatial distribution of \arii\ and \neii\ is shown in
Figure 2.  The morphologies are similar, thought not identical, and they
are also similar to the Pistol \HH\ region (Yusef-Zadeh and Morris, 1987,
see also Figer et al.~1998).  Energy distributions of the northern,
central, and southern parts of the Pistol Nebula are shown in Figure 3,
together with the energy distributions of four of the cocoon stars.
Photometry of these stars was obtained by measuring the flux within a 3
arcsec aperture and correcting for the aperture size using measurements of
single stars.  The Pistol Nebula spectra were extracted using $7\times
17\,$pixel synthetic aperture, and were normalised by the number of pixels.
The background was determined from the southernmost region of the field,
and subtracted.  It is clear from the energy distributions that the
silicate absorption is much deeper in the cocoon stars than in the Pistol
Nebula.  Table 1 lists the measured optical depths, $\tausil$, relative to
a power-law continuum fitted through the spectral points at 8 and $13\mic$.
For comparison, from the Lutz et al.~(1996) SWS spectrum of the Galactic
Centre we derive $\tausil = 2.0$.  The difference in the depth of the
feature in the Quintuplet Stars and in the Pistol is consistent with the
results of Nagata et al.~(1993).

\begin{table}[h] \begin{center}
\caption{Silicate feature optical depths}
\begin{tabular}{lc} \hline \hline 
Source~~~~~~~~~~~~~~~~~~~~~~~~~~~~~~~~~~~~& $\tausil$ \\ \hline
Q1 & 2.60 \\ Q2 & 2.51 \\ Q3 & 2.38 \\ Q9 & 2.52 \\ \hline 
north & 1.40 \\ centre & 1.32 \\ south & 1.08 \\ \hline \hline
\end{tabular}\end{center}\end{table}

\begin{figure}[bht]  \begin{center} \leavevmode
\centerline{\epsfig{file=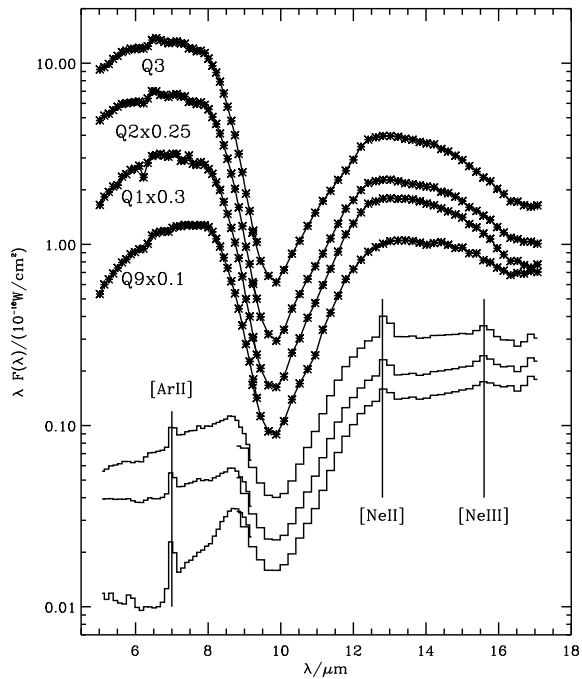,width=8cm}} \end{center}
\caption{\em Spectral energy distributions of the northern, central, and
southern part of the Pistol Nebula, shown with arbitrary normalisation, and
of the cocoon stars. }
\end{figure}

An SWS spectrum covering the 2.3---$47\mic$ range was obtained with the
Pistol Star centered in the aperture.  The AOT1-speed 4 mode was used, for
a total integration time of $\sim 2\,$hrs.  Pipeline-processed data was
cleaned by removing data from noisy detectors, normalising each scan to the
average of the complete dataset, and rebinning to a resolution of 1,000.
This result was then defringed by finding and removing the strongest
sinusoidal components.  The final spectrum is shown in Figure 4; it shows a
wealth of fine structure lines, many hydrogen recombination lines, and the
standard absorption features of interstellar origin (silicates at $9.7\mic$
and $18\mic$, water ice at $2.9\mic$, hydrocarbons at $3.4\mic$, and CO$_2$
ice at $4.3\mic$, though the latter are not obvious in the representation
given in Figure 4, which emphasises the emission lines.  Preliminary
analysis indicates that at $\lambda \lapp 5\mic$ the continuum and the
lines come primarily from the Pistol Star and its wind, while at longer
wavelength they come primarily from the hot dust and the ionised gas in the
\HH\ region.  A detailed analysis of the lines is in progress and will be
presented in a future paper.

\begin{figure*}[t]  \begin{center} \leavevmode
\centerline{\epsfig{file=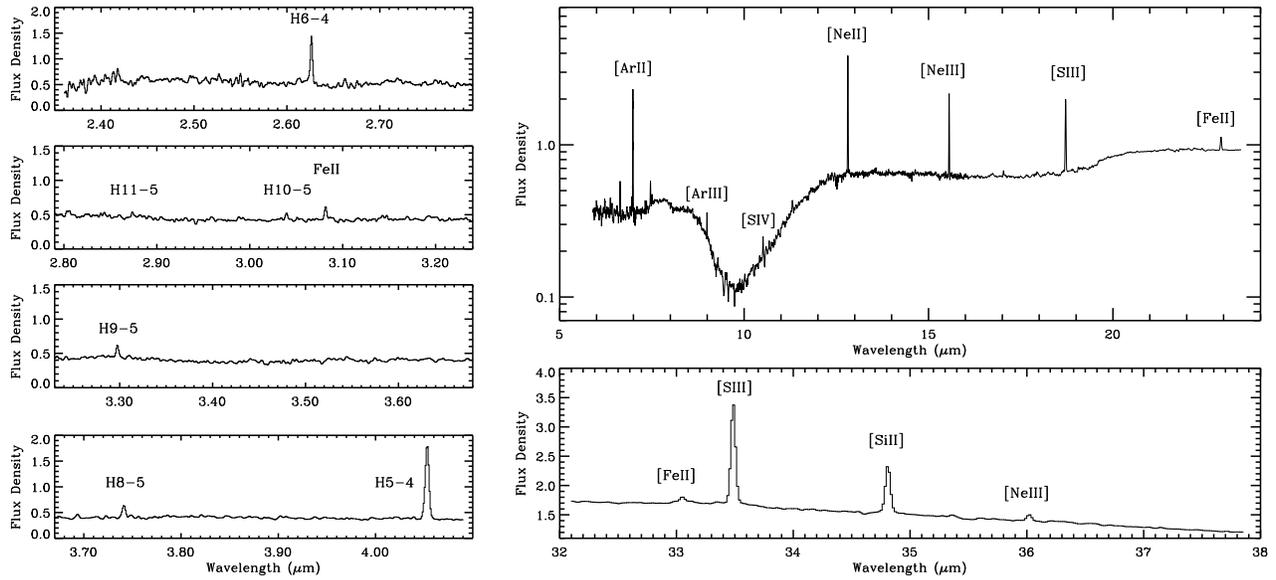,width=17cm}} \end{center}
\caption{\em SWS spectrum of the Pistol Nebula.  Different parts of the
spectrum are plotted separately for clarity.  The flux density is in units
of $10^{-16}\fu$.}
\end{figure*}

\section{DISCUSSION} 

The discovery of a spherical nebula around the Pistol Star is the main
result of this work presented at this workshop: our observations in the
thermal IR have revealed the full shell of ejecta symmetrically distributed
around the Pistol Star.  Up until now, only the ionized part of the Pistol
Nebula had been known, originally from radio continuum observations
(Yusef-Zadeh and Morris, 1987), but subsequently also from near IR
spectroscopy (see the Introduction).  Many LBVs are known to be surrounded
by shells of ejecta which is seen in reflection in the optical and near
infrared (Nota et al.~1995) and via its thermal dust emission in the mid
infrared (Trams et al.~1996, 1999).  The size of the Pistol Nebula is
comparable to that of other LBVs (Nota et al.~1995, Table 1), though unlike
other LBVs, the nebular structure appears more as a full sphere than as a
shell.  Since the ejecta is not uniformly ionised, the ionisation probably
does not come from the Pistol Star, which consequently must be relatively
cool.  This argues in favour of the cooler and lower luminosity class of
models proposed by Figer et al.~1998.  More likely, the ionisation come
from other hot stars in the Quintuplet Cluster (FMM95, FMM99), of which
there are several north of the ionised ridge.  Other cluster stars (no.~76,
151, and 157 in FMM99) are located over the Pistol Nebula in projection,
and two of these (no.~7 and 157) may be physically inside the nebula and
could be responsible for the details of the structures seen in the ionised
line images.

The relative strength of the high excitation lines in the SWS spectrum also
indicates that the ionisation must come from stars stars much hotter than
the Pistol Star.  The ratios of \siv/\siii, of \ariii/\arii, and of
\neiii/\neii\ in the Pistol spectrum are much stronger than in the Galactic
Centre spectrum obtained by Lutz et al.~(1996), from which these authors
derive an average effective temperature of the ionising sources of 35,000
K.  We deduce that at least some of the stars ionising the Pistol material
must be hotter than 35,000 K, and the Wolf-Rayet stars in the Quintuplet
Cluster are clearly indicated as the ionising stars.

The above scenario assumes that the Pistol Star and the cocoon stars are
all members of the Quintuplet Cluster, though strictly speaking, this
scenario involves only the Pistol Star and the hot stars in the cluster,
and not the cocoon stars.  In this context, the low silicate absorption
optical depth of the the Pistol Nebula compared to the cocoon stars is
puzzling.  Either there is a steep gradient in extinction over the cluster,
decreasing rapidly from north to south, or the extinction is spatially
uniform, in which case the cocoon stars and the Pistol Star are at
different distances along the line of sight.  Alternatively, the silicate
absorption of the cocoon stars is not entirely interstellar, but it is
partly intrinsic.  Figer et al.~(1998) reviewed extinction estimates to the
Quintuplet Cluster.  They estimate $\Ak = 3.28$ and adopt the Rieke, Rieke
and Paul (1989) reddening law as the one that best characterises the
reddening towards the hot stars in the Quintuplet.  This gives $\Av = 29\pm
2$ to the hot stars, and is an appropriate value for the Pistol Star,
placing the whole group physically close to the Galactic Centre.  Using
$\Av = 29$, we obtain $\Av/\tausil = 20.5$ for the Pistol Nebula, a value
comparable to that found by Roche and Aitken (1984) for nearby Wolf-Rayet
stars and B supergiants.  At the same time, the same $\Av$ gives
$\Av/\tausil = 11.6$ for the cocoon stars, a value more typical of sources
in the Galactic Centre (Rieke and Lebofsky 1985).  We do not offer a
solution to this problem at this time.

In conclusion, we have solved one dilemma by discovering the full shell of
ejecta around the Pistol Star, thus further confirming its LBV nature,   but
we have opened a new one by finding that $\tausil$ varies drastically
between the Pistol Star/Nebula and the cocoon stars.   These are only the
first results from this data set;  further analysis is in progress, and in
an upcoming paper we will present a detailed analysis of the emission
lines, which we will use to put constraints on the nature of the ionising
radiation and of the extinction.


\end{document}